\documentclass{article}
\usepackage{spconf,amsmath,graphicx}
\usepackage{cite}
\usepackage{graphicx}
\usepackage{booktabs}
\usepackage{float}
\usepackage{enumitem}
\usepackage{multicol}
\usepackage{tikz}
\usepackage{hyperref}

\usepackage[normalem]{ulem}
\useunder{\uline}{\ul}{}
\usepackage{subcaption}

\title{ Frame-type sensitive RDO Control for Content-Adaptive Encoding}
\name{Vibhoothi, François Pitié,  Anil Kokaram \thanks{This work was funded by DTIF EI Grant No DT-2019-0068 and The ADAPT SFI Research Center \newline This paper is accepted in ICIP 2022,  \copyright IEEE 2022}}
\address{Sigmedia Group, \\
Department of Electronic and Electrical Engineering,\\
Trinity College Dublin,\\
Dublin, Ireland. \\
\{vibhoothi, pitief, anil.kokaram\}@tcd.ie}

\begin{document}
%
\maketitle
\begin{abstract}
Video transcoding is an increasingly important application in the streaming media industry. It has become important to investigate the optimisation of transcoder parameters for a single clip simply because of the immense number of playbacks for popular clips. In this paper, we explore the use of a canned optimiser to estimate the optimal Rate-Distortion (RD) tradeoff achievable for a particular clip. We show that by adjusting the Lagrange multiplier in RD optimisation on keyframes alone we can achieve more than 10$\times$ the previous BD-Rate gains possible without affecting quality for any operating point.
\end{abstract}
\section{Introduction}
\label{sec:intro}
In recent years, the growth in delivery of video at scale for applications like Streaming \& Broadcast of content (from Netflix, YouTube, Disney etc.) has inspired further research into content-adaptive Transcoding. The perennial goal of this area is to deliver high-quality content at increasingly lower bitrates by adapting the transcoder to the content presented, at a more fine-grained level of control. In 2013, YouTube was the first to adopt this strategy for its User-Generated-Content (UGC) content by building a pipeline that is based on clip popularity by re-processing a clip using an enhancement pre-processor combined with different baked-in transcoder settings. Around the same time, Netflix's by now seminal work on per-clip and per-shot encoding~\cite{netflix} showed that an exhaustive search of the transcoder parameter space can lead to massive gains in RD tradeoffs for a particular clip. Those gains easily compensate for the large once-off computational cost of transcoding because that clip may be streamed millions of times across many different CDNs thus saving bandwidth and network resources in general. That idea has since been refined into a much more efficient search process by applying the Viterbi algorithm across shots and parameter spaces~\cite{katsavounidis2018video}. In those efforts, the authors concentrate on optimisation of a high-level parameter (target bitrate or quantisation factor or objective quality) to generate the best bitrate ladder for a clip as part of a DASH stream. Recently we showed in~\cite{pcs2021ringis} that the
rate-distortion tradeoff could be directly addressed by applying a numerical optimisation scheme to estimate the appropriate Lagrange multiplier for a particular clip (see Section~\ref{section:background}). We observed an average BD-Rate improvement of
1.9\% for HEVC and 1.3\% for VP9~\cite{SPIERingis}.

In this paper, we propose that a content-adaptive transcoder optimisation at a deeper level within the encoder could result in higher gains. The  Lagrange multiplier is indeed used in different sub-modules of the codec and we propose to target particular modes instead of using a constant value for all the decision making in the codec.

Our contribution is thus to adjust our per-clip optimisation strategy and modulate the Lagrange multiplier for 1) the different frame types in HEVC and AV1 and 2) the image partitioning process of HEVC and AV1 (see Section~\ref{section:key-opt}).

Our experiments in Section~\ref{section:expt-setup} show that such a targeted optimisation of the Lagrange multiplier can lead to average BD-Rate (MS-SSIM) gains of 4.9\%, 1.5\%, and best gains of 29.2\% and 5.6\% for AV1 and HEVC respectively.


\section{Background}
\label{section:background}
The work of Sullivan et al~\cite{sullivan1998rate} laid the foundations for optimising the rate-distortion tradeoff in modern video codecs. By taking a Lagrange multiplier approach the joint optimisation problem is posed as the minimisation of $J = D + \lambda R$, where $\lambda$ is the Lagrange multiplier controlling the tradeoff. This idea is the basis for the rate-distortion-optimisation (RDO) process used especially in making mode decisions in modern codecs. The independent variable in this optimisation is usually $Q$, a quantiser step size. Increasing $Q$ reduces rate $R$ but increases distortion $D$. A judicious choice of $\lambda$ yields different $R,D$ pairs resulting from the minimisation problem. By optimising over a test video clip set, $\lambda$ is related to $Q$ through an empirical relationship. For example, in AV1~\cite{av1paper} $\lambda$ is estimated from $q_i$ (the quantiser parameter)  as follows:
\begin{equation}
  \lambda = q_{dc} ^2 \times (A + 0.0035 \times q_{i}), \label{lambdaeq:av1}
\end{equation}
where $A$ is a constant depending on frame type ($3.2 \leq A \leq 3.3$) and $q_{dc}=f(q_i, A)$ is another defined mapping implemented with a discrete valued LUT. In HEVC~\cite{hevcoverview} the relationship is simpler (\(\lambda = 0.57 \times 2^{\frac{q_i-12}{3}}\)).

The range of $Q$ is also different for different codecs, $[0,51]$ for HEVC and $[0,63]$ for AV1.

It is clear that this $\lambda-Q$ relationship is not necessarily optimal and that, ideally, $\lambda$ should be content dependent. Several authors have explored the idea of adapting $\lambda$ to features measured from the content. For example, Zhang and Bull~\cite{zhangbulllambdahevc, zhangbulllambdahevcv2} altered  $\lambda$ in each frame based on the distortion per frame. In our previous work~\cite{EIRingis,SPIERingis, pcs2021ringis}, we introduced the idea of an adaptive $\lambda$ on a clip basis using a single modified $\lambda = k\lambda_o$ across all the frames in a clip. Here $\lambda_o$ is default value from the standard empirical relationship e.g. Eq.~(\ref{lambdaeq:av1}). The motivation was to explore exactly how much gain was possible using this idea. We deployed simple numerical minimisers (Brent's method, Golden-search~\cite{numericalmethods}) to directly address estimation of $\lambda$ by using the BD-Rate itself as the cost function for optimisation. Later work considered the use of ML techniques to reduce the computation required by predicting that optimum from low-cost features~\cite{pcs2021ringis}.   We observed an average improvement of 1.87\%, 1.324\% for HEVC and VP9 respectively, with gains up to 23\% on certain clips, and gains of more than 1\% on 20\% of clips on the YouTube-UGC dataset~\cite{SPIERingis}.



\section{Keyframe Optimisation}
\label{section:key-opt}

In Section~\ref{section:results} we show that a direct optimisation of $\lambda$ for AV1 and HEVC, only yields an average gain of 0.539\% and 0.097\%. Our idea here is that greater gains can be made by optimising $k$ for particular modes instead of using a constant $k$ for all the decision making in the codec.
%
%
RDO theory tells us that $\lambda$ should be constant over the considered video segment (eg. DASH chunk). By establishing a fixed empirical $\lambda-Q$ relationship, codecs break that theory, and we note that the definition of $\lambda$ in AV1, through the constant $A$ in Eq.~(\ref{lambdaeq:av1}), actually depends on the frame-type. We propose here to optimise this further.

Over our selected dataset of 10 sequences (1300 Frames), reference frames (generically called keyframes here) in HEVC and AV1 are more than $5$ to $10$ times larger w.r.t. bits than other frames.  Hence we target $\lambda$ optimisation for keyframes in HEVC and AV1. The reference frame management of HEVC, has not changed much from H.264/AVC with I-Frames, B-Frames and P-Frames being used. Each B-Frame employs 2-5 reference frames depending on a measure of frame importance.
In contrast, AV1 employs up to 8 reference frames~\cite{av1:adaptivepred}. Here we consider three main keyframe types: the usual reference intra-coded frame {\tt KEYFRAME}, an alternate reference frame {\tt ARF\_FRAME} used in prediction but does not appear in the display, and an intra coded frame encoded with higher quality {\tt GOLDEN\_FRAME}.

\begin{table}
\centering
\small
\begin{tabular}{ll}
\toprule
\textbf{Codec} & \textbf{Targeted Frame-Types for Optimisation} \\ \midrule
\textbf{AV1} & 
1. All Frames     \\ 
& 2. Key-Frames (KF)   \\ 
& 3. Golden-Frames (GF), Alt-Ref (ARF)  \\
& 4. KF, GF, ARF     \\ \midrule
\textbf{HEVC} & 
1. All Frames     \\ 
& 2. I-Frames   \\ 
& 3. B-Frames \\ \bottomrule
\end{tabular}
\caption{The different frame-types targeted for specific Lagrange multiplier optimisation in our experiments.}
\label{table:searchmodes}
\end{table}


 Table~\ref{table:searchmodes} shows the different keyframe optimisation groupings we have identified. {\em All Frames} refers to the default optimisation case in which $k$ is the same for all frame types. In each other row, we indicate the keyframe types for which $k \neq 1$, i.e. in these cases we keep $\lambda = \lambda_o$ for all frame types and optimise $\lambda$ only for the grouping shown where $\lambda = k\lambda_0$.

In the keyframe optimisation modes discussed, all the RD decisions within that keyframe employ the same  $k$. We denote this as {\em Top} level optimisation. However, both HEVC and AV1 employ an adaptive partitioning of the frame for compression~\cite{hevc:texturebase,av1:adaptivepred}. The idea is that in textureless regions, large blocks can be compressed using a single DCT while in textured regions smaller blocks are preferred. The partitioning tessellation is estimated based on RD optimisation and hence is also likely to have an important impact on the compression of a keyframe. Thus we also investigate the impact of tuning $\lambda$ for the partitioning optimisation alone within the keyframe groupings. We denote this as {\em Partition} level optimisation.

In the optimisation process itself we select $k$ to maximise the BD-Rate gain for each clip ($m$) using MS-SSIM~\cite{msssimpaper} as the quality metric ($D_m$) for the $m^{\text{th}}$ clip. The cost function $C_m(k)$  is therefore as follows:
\begin{align}
    C_m(k) & = \text{BD-Rate}(R_m(k\lambda_o, D),R_m(\lambda_o, D)) \nonumber \\
    & \propto \int_{D_1}^{D_2} \left(R_m( k\lambda_o, D)-R_m(\lambda_o, D\right) dD,
\end{align}
where $R_m( k\lambda_o, D)$ is the bitrate of the $m^{\text{th}}$ clip at quality $D$ and using ${\bf{\lambda}} = k\lambda_o$. That rate measurement is derived from the MS-SSIM RD characteristic generated using $N$ QP measurements. The range $D_1, D_2$ is as defined for the BD-Rate calculations~\cite{BDRate}. We use Brent's optimiser~\cite{numericalmethods} for minimisation of the cost function $C_m(k)$.

\begin{table*}
\resizebox{\textwidth}{!}{%
\begin{tabular}{@{}lllrrrrrrrrr@{}}
\toprule
Codec & 
\multicolumn{1}{l}{\begin{tabular}[c]{@{}l@{}}$\lambda$ Modified at\end{tabular}} & 
Frame-Type &
\multicolumn{1}{l}{\begin{tabular}[c]{@{}l@{}}Avg. $\hat{k}$ \\ Value\end{tabular}} & \multicolumn{1}{l}{\begin{tabular}[c]{@{}l@{}}Avg.\\ BDR(\%)\end{tabular}} &  \multicolumn{1}{l}{\begin{tabular}[c]{@{}l@{}}Max.\\ BDR(\%)\end{tabular}} & \multicolumn{1}{l}{\begin{tabular}[c]{@{}l@{}}Min.\\ BDR(\%)\end{tabular}} & \multicolumn{1}{l}{\begin{tabular}[c]{@{}l@{}}Avg. \\ Iters\end{tabular}} & \multicolumn{1}{l}{\begin{tabular}[c]{@{}l@{}}Avg. \\Bitrate \\ Savings(\%)\end{tabular}} & \multicolumn{1}{l}{\begin{tabular}[c]{@{}l@{}}Avg. RD2 \\ Bitrate \\ Savings(\%)\end{tabular}} & \multicolumn{1}{l}{\begin{tabular}[c]{@{}l@{}}Avg. \\MS-SSIM \\ Change (dB)\end{tabular}} & \multicolumn{1}{l}{\begin{tabular}[c]{@{}l@{}}Avg. \\VMAF \\ Change\end{tabular}} \\ \midrule
 &  & \textit{All Frames} & \textit{1.247} & \textit{-0.539} & \textit{-1.546} & \textit{0.261} & \textit{8.7} & \textit{-1.635} & \textit{-5.243} & \textit{0.327} & \textit{1.342} \\ 
{AV1}  & \textit{Top} & \textit{KF } & \textit{2.878} & \textit{-3.832} & \textit{-24.757} & \textit{-0.392} & \textit{9.1} & \textit{-3.458} & \textit{-8.575} & \textit{0.026} & \textit{0.075} \\ 
 &  & \textit{GF, ARF } & \textit{2.226} & \textit{-1.841} & \textit{-4.027} & \textit{0.152} & \textit{8.7} & \textit{-4.603} & \textit{-5.173} & \textit{0.097} & \textit{0.170} \\ 
{\ul \textbf{}} & {\ul } & {\ul \textit{KF, GF, ARF }} & {\ul \textit{2.494}} & {\ul \textit{-4.924}} & {\ul \textit{-29.159}} & {\ul \textit{-0.202}} & {\ul \textit{10}} & {\ul \textit{-5.817}} & {\ul \textit{-10.503}} & {\ul \textit{0.151}} & {\ul \textit{0.832} } \\ \midrule 
 &  & All Frames & 1.203 & -0.492 & -1.637 & 0.190 & 9.6 & -1.569 & -4.337 & 0.269 & 1.145 \\ 
{AV1}  & Partition & KF  & 2.748 & -3.692 & -23.763 & -0.381 & 9 & -3.318 & -8.394 & 0.023 & 0.082 \\ 
 &  & GF, ARF  & 2.149 & -1.754 & -3.897 & 0.149 & 8.9 & -4.306 & -4.761 & 0.085 & 0.137 \\ 
 &  & {\ul KF, GF, ARF } & {\ul 2.467}  & {\ul -4.772} & {\ul -29.049} & {\ul 0.000} & {\ul 9.5} & {\ul -4.618} & {\ul -8.761} & {\ul 0.127} & {\ul 0.708} \\  \midrule
 &  & \textit{All Frames} & \textit{0.964} & \textit{-0.097} & \textit{-0.323} & \textit{0.077} & \textit{9.7} & \textit{2.232} & \textit{1.520} & \textit{-0.045} & \textit{-0.210} \\ 
{HEVC} & \textit{Top} & {\ul \textit{I-frames }} & {\ul \textit{2.051}} & {\ul \textit{-1.458}} & {\ul \textit{-5.552}} & {\ul \textit{0.038}} & {\ul \textit{9.3}} & {\ul \textit{-3.260}} & {\ul \textit{-4.671}} & {\ul \textit{0.070}} & {\ul \textit{0.431}} \\ 
\textbf{} &  & \textit{B-frames } & \textit{0.867} & \textit{-0.682} & \textit{-4.271} & \textit{0.022} & \textit{7.5} & \textit{14.671} & \textit{4.228} & \textit{-0.124} & \textit{-0.704} \\ \midrule
 &  & All Frames & 0.942 & -0.143 & -0.625 & 0.078 & 7.9 & -0.726 & 0.933 & -0.031 & -0.104 \\ 
{HEVC} & Partition & {\ul I-frames } & {\ul 1.467} & {\ul -0.495} & {\ul -2.513} & {\ul 0.058} & {\ul 9.5} & {\ul 0.627} & {\ul -1.201} & {\ul 0.009} & {\ul 0.063} \\ 
 &  & B-frames & 0.954 & -0.339 & -1.806 & 0.042 & 9.5 & -2.803 & 1.440 & -0.059 & -0.314 \\ \bottomrule
\end{tabular}}%
\caption{Summary results for both AV1 and HEVC-HM. The underlined result is the best for each codec employing a particular $\lambda$ modification level. \textit{Note: negative values indicate the result is better}. Optimally modifying $\lambda$ at the keyframe level (I-Frames for HEVC and KF/GF/ARF for AV1) shows the most BD-Rate gains with negligible impact on quality. }
\label{table:mainresults}
\end{table*}

\section{Experimental Setup}
\label{section:expt-setup}
For our experiments, we designed a scalable framework for canned-optimisation of transcoder parameters by extending AreWeCompressedYet~\cite{awcy}. We deployed the stable releases for both AV1 ({\tt libaom-av1-3.2.0, 287164d}) and HEVC ({\tt HEVC-HM-16.24, fd452ecd}) with modifications to allow $k$ to propagate to the desired mode decision from a command line argument.

Our dataset consists of  5 clips from YouTube's User-Generated Clips (UGC) dataset (\textit{Set A})~\cite{wang2019youtube}, and 5 from AV2-A2-2K set  AOM-CTC (\textit{Set B})~\cite{aomctc}. The first 130 frames are used from these sources at $1920 \times 1080$ resolution and 4:2:0 color format.  The Random Access (RA) encoding mode is employed. This mode is commonly used for streaming as it allows users to randomly seek into any frame of the clip.

Configurations for the encoders were made as per Common Test Conditions (CTC) of AV1~\cite{aomctc} and HEVC~\cite{hevcctc} using constant QP and fixed QP offsets. The QP points for libaom-av1 were \{27, 39, 49, 59, 63\} and for HEVC-HM \{22, 27, 32, 37, 42\}. The Rate-Distortion points for the specified QPs are computed with help of  libvmaf\cite{libvmafurl}, a standard and open source video quality evaluation library used in industry.

Further implementation details, as well as an exhaustive analysis of our results can be found on our project page\footnote{\href{https://gitlab.com/mindfreeze/icip2022}{https://gitlab.com/mindfreeze/icip2022}}.

\section{Computational Load}
\label{section:comp-load}
The computational load of these experiments is significant. The calculation of our cost function requires first the RD curve generation for every one of our clips using $5$ QP points and $\lambda=\lambda_o$. That is $N=5$ encoder invocations for $M$ clips, i.e. $NM$ encoder invocations. Thus $P$ iterations of our optimise require $PNM$ encoder invocations. On average our optimiser took $P=9$ iterations to converge (see Table~\ref{table:mainresults}), and this implies an average of $45$ encoder invocations for 1 Clip. Even parallelising the RD curve generation using 5 threads, each for one operating point on the RD curve, it took an average of $329$ hours of single CPU Wall Clock for our M=10 clips in AV1 and $87$ hours in HEVC. We used 125 threads on a 2nd Gen AMD EPYC 7720p based multiprocessing system. Because of this load, we do not intend that this optimisation process can be used in a production environment but instead it shows the gains possible in adapting $\lambda$.


\begin{figure}
    \centering
    \includegraphics[width=0.95\linewidth]{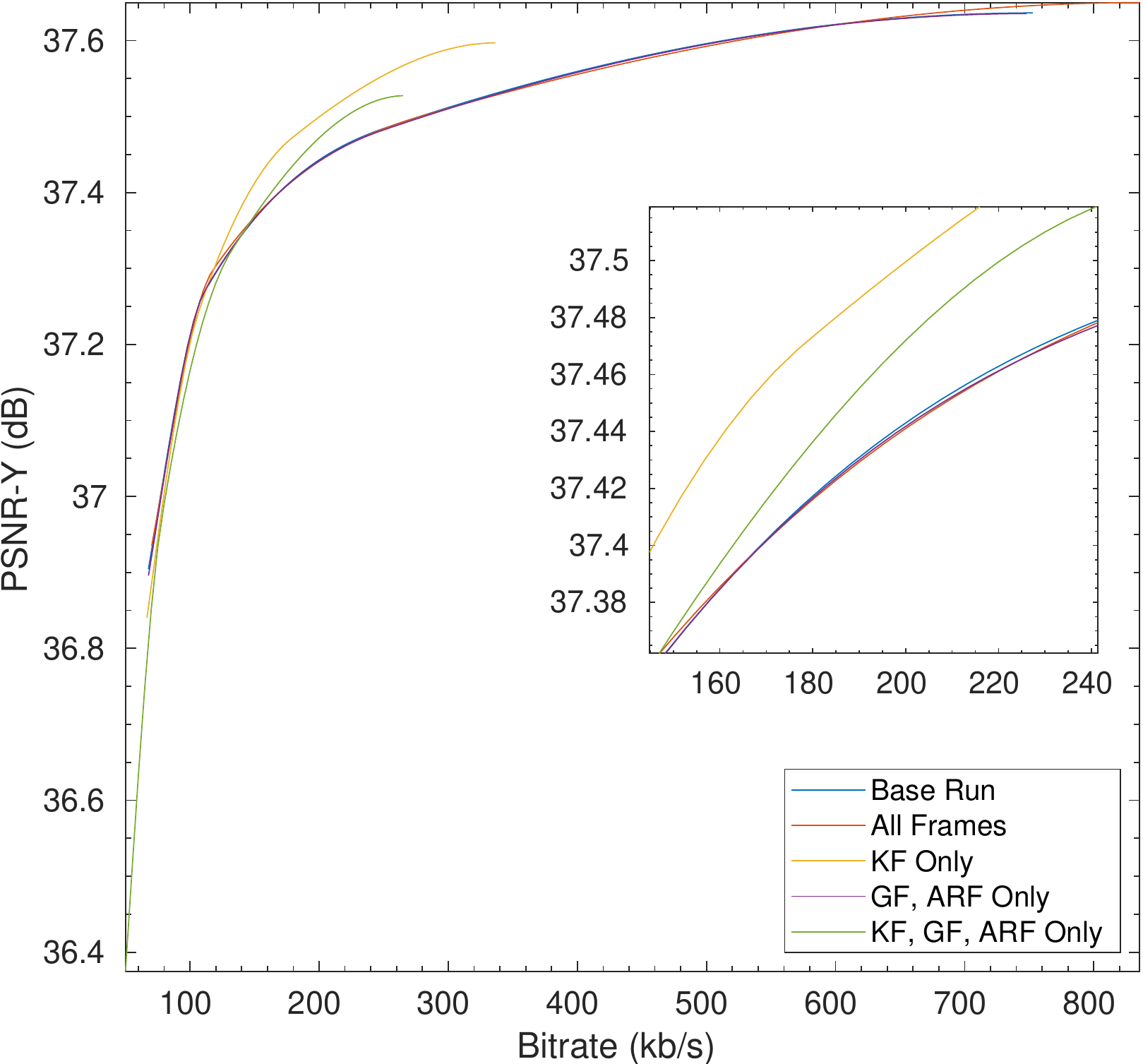}
    \caption{AV1 RD curves for {\em DinnerSceneCropped}~\cite{aomctc}  with different keyframe optimisations. In tuning for the {\em KF, GF, ARF} frame group, the curve has shifted to the left, resulting in mean bitrate savings of 24\% (at QP39 64.6\%). The inset emphasises the gain especially at low bitrates.}
    \label{fig:av1_rdcurve}
\end{figure}

\section{Results}
\label{section:results}
Table~\ref{table:mainresults} shows our measured BD-Rate (MS-SSIM), MS-SSIM (dB), VMAF~\cite{libvmafurl ,vmafpaper} gains using various summary statistics. We present the experiments organised by codec, $\lambda$ modification level and frame-type.  The underlined values are the best gains for each codec and $\lambda$ modification level.

The {\em Avg $k$} is $\neq 1$ in all cases, hence verifying our claim that a better $\lambda$ exists. Also impact on quality (MS-SSIM and VMAF) is negligible showing that gains in bitrate are not offset by loss in quality. Significantly, the BD-Rate gains achievable using keyframes are as much as $10\times$ and $15\times$ better (AV1, HEVC) than using the same $k$ for {\em All Frames} types.  Overall, the table shows that there is more to gain with this optimisation strategy for AV1 than HEVC ($\approx 3 \times$)

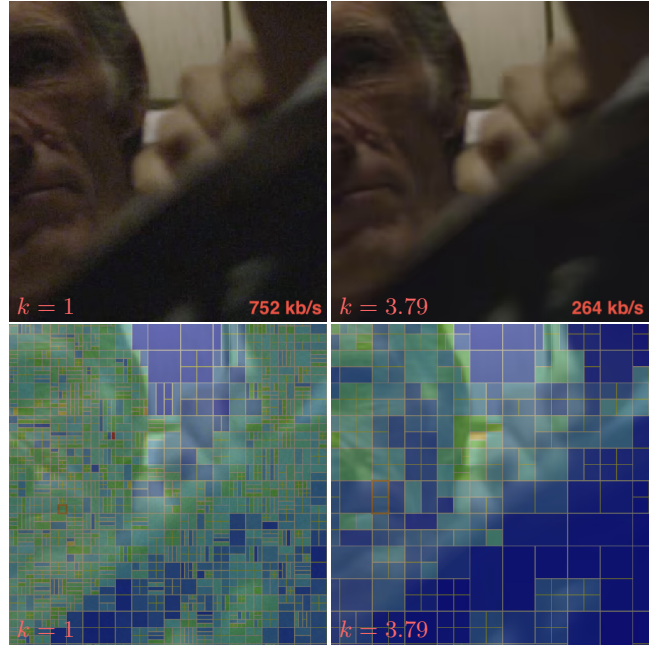
\begin{figure}
    \centering
    \setlength\tabcolsep{1pt}
    \renewcommand{\arraystretch}{0.4}
    \begin{tabular}{ll}
    \begin{tikzpicture}
    \node[anchor=south west,inner sep=0] (image) at (0,0) {\includegraphics[width=0.49\linewidth]{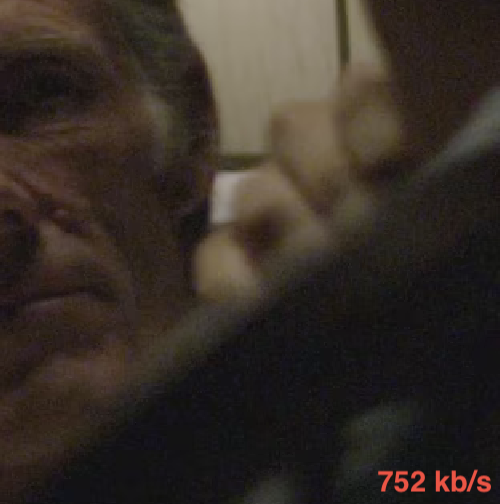}}; \begin{scope}
    \node[draw=none, anchor=south west, text=red!60!white, inner sep=0, font=\sffamily] at (0.1,0.1){\small $k=1$};
    \end{scope}\end{tikzpicture}
    & 
    \begin{tikzpicture}
    \node[anchor=south west,inner sep=0] (image) at (0,0) {\includegraphics[width=0.49\linewidth]{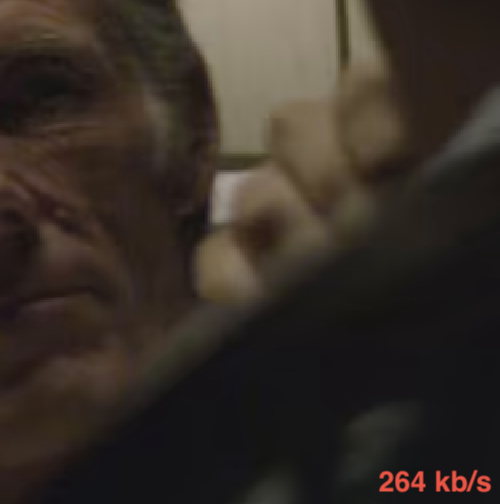}}; \begin{scope}
    \node[draw=none, anchor=south west, text=red!60!white, inner sep=0, font=\sffamily] at (0.1,0.1){\small $k=3.79$};
    \end{scope}\end{tikzpicture}
\\
    \begin{tikzpicture}
    \node[anchor=south west,inner sep=0] (image) at (0,0) {\includegraphics[width=0.49\linewidth]{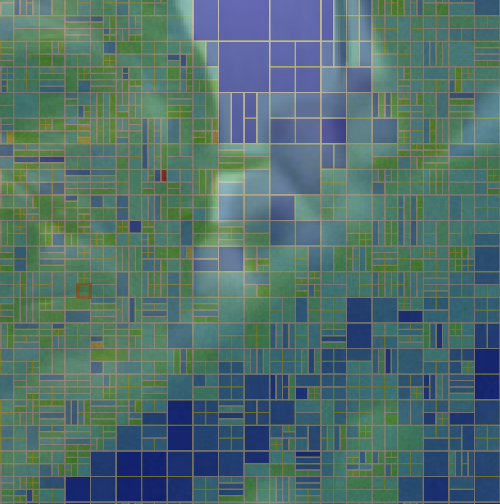}}; \begin{scope}
    \node[draw=none, anchor=south west, text=red!60!white, inner sep=0, font=\sffamily] at (0.1,0.1){\small $k=1$};
    \end{scope}\end{tikzpicture} &
        \begin{tikzpicture}
    \node[anchor=south west,inner sep=0] (image) at (0,0) {\includegraphics[width=0.49\linewidth]{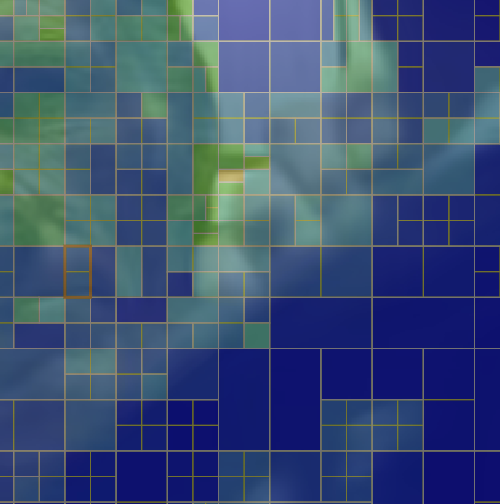}}; \begin{scope}
    \node[draw=none, anchor=south west, text=red!60!white, inner sep=0, font=\sffamily] at (0.1,0.1){\small $k=3.79$};
    \end{scope}\end{tikzpicture}
    \end{tabular}
    \caption{Crop of keyframe 1 from \textit{DinnerSceneCropped}. Left: default encoding ($k=1$), Right: proposed optimisation ($\hat{k}=3.79$). The top row shows that the optimised output has less noise and uses 65\% less bits. The bottom row shows the tiling pattern and bit allocation with a colour heatmap (blue = less bits and green/yellow/red = increased bit allocation). That row shows the gains are due to a combination of increased smoothing leading to a less fragmented tiling.}
    \label{fig:blockandbitav1}
\end{figure}

Considering {\em AV1, Top} we see that using the same $k$ for the 3 reference frame group {\em KF, GF, ARF} give the best performance with a BD-Rate gain of 4.9\% but that is just 1 \% more than using adaptation on {\em KF} alone, implying that optimisation on just {\em KF} gives the most impact. Comparing {\em AV1, Top} with {\em AV1, Partition} we see that optimising $k$ purely for the partition decision for {\em KF, GF, ARF} frame types yields 4.8\% gain, which is more than 98\% of the gain using top level modification (4.9\%). That means that it is the partition RDO process for reference frames that is benefiting the most from optimisation of $k$.
This is well illustrated in Figure~\ref{fig:blockandbitav1} showing a portion of a keyframe under AV1 encoding for the best performing clip. The optimal $k$ has caused a radical change in partitioning itself caused by an increased noise reduction effect. The removal of the noise in textureless regions causes those regions to be better tiled with larger blocks.

In HEVC the situation is slightly different. The best average gains (1.5\%) correspond to {\em I-frame} optimisation but partitioning accounts for only about 30\% of that gain. Hence other RDO decisions are having equal impact.

To examine the bitrate savings at more typical streaming rates (4-5 Mbps) we report on the average savings at the second operating point from our QP range which achieves roughly these rates. In AV1 this is 39 and 27 in HEVC. This is shown in column RD2 of Table~\ref{table:mainresults} and indicates that the bitrate savings at this operating point are more than $2\times$ that across all operating points (column to the left of RD2).  The trend across AV1 and HEVC is the same at this operating point with more bitrate savings to be gained in AV1 than HEVC. 


The best gain for a single clip is significant (shown in column {\em Max BDR}): about 30\% for AV1 and 6\% for HEVC. 
Figure~\ref{fig:av1_rdcurve} shows the piecewise cubic Hermite interpolating polynomial (PCHIP) interpolation~\cite{pchippaper} of the AV1 RD Points used in the DinnerSceneCropped clip which gave these maximum BD-Rate gains at average bitrate savings of 24\% (at QP39 it was 64.67\%). The plot shows the dramatic shift of the RD curve to the left when $\lambda$ optimisation is used. The same clip showed maximum gains with HEVC but at lower levels e.g. average bitrate savings of 14.8\% (at QP27 it was 26\%) and hence we do not show that RD characteristic.

\section{Conclusions}
\label{section:conclusion}
We have presented a new method of improving the existing system of per-clip
optimisations by modulating $\lambda$ based on frame-types. The method yields
average BD-Rate improvements of 4.9\% for AV1 and 1.5\% for HEVC, with the
selective optimisation of reference frames contributing to about 90\% of these
improvements. We also showed that the improved tiling was responsible for most
of the improvements inside the AV1 encoder but only accounts for 30\% of the
improvements in HEVC. Future work will consider reducing the computation cost by
exploiting proxy resolutions.

\bibliographystyle{IEEEbib}
\bibliography{main.bib}

\end{document}